\documentclass[aps,pre,preprint,showpacs,amsmath,amssymb,amsfonts,lengthcheck,longbibliography,superscriptaddress]{revtex4-2}

\usepackage{changes} 
\usepackage{ulem} 
\def\dbar{{\mkern3mu\mathchar'26\mkern-12mu d}}
 \usepackage[utf8]{inputenc}
\usepackage{graphicx}
 \usepackage{subfigure}
\usepackage{amsthm}
\usepackage{mathtools}
\usepackage{amsmath}
\usepackage{verbatim}
\usepackage{dcolumn}
\usepackage{bm}

\usepackage{color}
\usepackage[colorlinks=true,linkcolor=blue,urlcolor=blue,citecolor=blue,pdfusetitle]{hyperref}

\usepackage{xcolor}
\usepackage{physics}
\usepackage{dsfont}
\usepackage{orcidlink}
\usepackage{longtable}
 \usepackage{xargs}                      

\NewDocumentCommand\eq{ms}{%
  \IfBooleanTF{#2}{\begin{align}#1\end{align}}{\begin{align}
#1
\end{align}}
}
\newcolumntype{M}[1]{>{\centering\arraybackslash}m{#1}}
\definecolor{rojo}{RGB}{90, 30, 175}

\begin{document}
 
\title{Quantum Level-Crossing Induced by Anisotropy in Spin-1 Heisenberg Dimers: Applications to Quantum Stirling Engines}
 
\author{Bastian Castorene\,\orcidlink{0009-0002-9075-5716}}
\email{bastian.castorene.c@mail.pucv.cl}
\affiliation{Instituto de Física, Pontificia Universidad Católica de Valparaíso, Casilla 4950, 2373223 Valparaíso, Chile}
\affiliation{Departamento de Física, Universidad Técnica Federico Santa María, 2390123 Valparaíso, Chile}

\author{Vinicius Gomes de Paula\,\orcidlink{0000-0002-6859-2917}}
\affiliation{Institute of Physics, Fluminense Federal University, Av. Gal. Milton Tavares de Souza s/n, 24210-346 Niterói, Rio de Janeiro, Brazil}

\author{Francisco J. Peña\,\orcidlink{0000-0002-7432-0707}}
\affiliation{Departamento de Física, Universidad Técnica Federico Santa María, 2390123 Valparaíso, Chile}

\author{Clebson Cruz\,\orcidlink{0000-0003-3318-1111}}
\affiliation{Quantum Information and Statistical Physics Group, Western Bahia Federal University - Campus Reitor Edgard Santos, Bertioga Street 892, Morada Nobre I, 47810-059 Barreiras, Bahia, Brazil}

\author{Mario Reis\,\orcidlink{0000-0002-3349-3003}}
\affiliation{Institute of Physics, Fluminense Federal University, Av. Gal. Milton Tavares de Souza s/n, 24210-346 Niterói, Rio de Janeiro, Brazil}

\author{Patricio Vargas\,\orcidlink{0000-0001-9235-9747}}
\affiliation{Departamento de Física, Universidad Técnica Federico Santa María, 2390123 Valparaíso, Chile}

\date{\today}

\begin{abstract}
  
This work explores the thermodynamic performance of a quantum Stirling heat engine implemented with an anisotropic spin-1 Heisenberg dimer as the working medium. Using the Hamiltonian of the system, we analyze the interplay of anisotropy, magnetic field, and exchange interactions and their influence on the energy spectrum and the quantum level crossing. Our results reveal that double-degenerate point (DDP) and a triple-degenerate point (TDP) play pivotal roles in shaping the operational regimes and efficiency of the quantum Stirling engine. At those points, the Carnot efficiency reaches higher work output and enhanced stability, making it a robust candidate for optimal thermodynamic performance. These findings highlight the potential of anisotropic spin systems as viable platforms for quantum heat engines and contribute to advancing the field of quantum thermodynamics.

\end{abstract}

 \maketitle

\section{Introduction}

The development of quantum thermodynamics has provided a deeper understanding of the fundamental principles governing the energy transfer processes at the quantum scale \cite{alicki1979quantum,Quan:07,kumar2023introduction,scovil1959three,solfanelli2020nonadiabatic, myers2021quantum,kosloff1984quantum,binder2018thermodynamics,goold2016role,vinjanampathy2016quantum,landi2021irreversible,10.1116/5.0083192}. In particular, the study of quantum heat engines has become a prominent research field, revealing novel phenomena that emerge when thermodynamic systems are analyzed under the rules of quantum mechanics. Among these quantum heat machines, the Stirling cycle is a versatile framework for studying energy conversion processes in quantum systems \cite{rossnagel2014nanoscale,wu2006performance,lin2003quantum,abah2022stirling,feldmann1996quantum,geva1992quantum,huang2014quantum, ghannadan2021magnetic, vargova2021unconventional, cruz2022, araya2023magnetic,PhysRevE.110.044135,Castorene2,Rojas2024,pili2023quantum}. 

In this scenario, spin systems have been widely employed as paradigmatic models in quantum thermodynamics due to their rich energy structures and experimental feasibility in platforms such as nuclear magnetic resonance, trapped ions, and superconducting qubits \cite{nmr_spin, trapped_ions, superconducting_qubits,PhysRevE.110.044135,e22070755,Rojas2024,makarova2017acs}. In particular, the anisotropic spin-1 Heisenberg dimer provides a unique platform to analyze thermodynamic properties and quantum level-crossing through the interplay of anisotropy and spin-spin interactions \cite{ghannadan2021magnetic,strevcka2005magnetic,STRECKA20083146,belle2001,biswas2009,heisenberg_model1, heisenberg_model2,QE1, heisenberg_model3, PhysRevE.110.044135, pili2023quantum,depaula2024}. By systematically varying the anisotropy and external magnetic field, we uncover how these parameters shape the energy spectrum and influence the operational efficiency of the Stirling engine. Such behavior is intricately linked to the underlying quantum critical points (QCPs) and quantum phase transitions, phenomena that have been extensively studied across diverse quantum spin systems \cite{qcp1,qcp2,qcp3,qcp4,qcp5,qcp6,qcp7,qcp8,qcp9,qcp10,qcp11,qcp12,qcp13,qcp14,qcp15,qcp16,qcp17,qcp18,qcp19,newanisotropy1,WOS:001493024900001,WOS:001495458500001}. 

Specifically, numerous methods have been developed to quantify entanglement in spin-1 systems, such as spin-spin correlations, entanglement entropy, coherence, negativity and quantum discord \cite{ghannadan2021magnetic,QD1,QD2,QD3,QD4,QD5,QD6,QD7}. For spin-1/2 particles, a widely used entanglement measure is concurrence, introduced by Wootters \cite{Wootters1,Wootters2,Wootters3,Wootters4}, which has become an established tool for quantifying the entanglement between spin-1/2 particle pairs. However, due to its original formulation, bipartite concurrence is not applicable to particles with different spin magnitudes.

In the context of quantum thermal machines, quantum critical points in thermal entanglement for spin-1/2 systems have been shown to indicate regions where the system achieves maximum classical efficiency \cite{PhysRevE.110.044135,araya2023magnetic,PURKAIT2022128180,Peña1,Peña2,PhysRevLett.109.203006,Peña3,Peña4,newanisotropy1,newanisotropy2,newanisotropy3,newanisotropy5,clebson2,clebson3,clebson4,clebson5,clebson6,clebson7,clebson8,clebson9}. To address the limitations of concurrence for higher-dimensional systems, generalized concurrence measures have been introduced \cite{GConcu1,GConcu2,GConcu3,GConcu4,GConcu6,GConcu7,Gconcu8,GConcu9}. In previous studies, generalized concurrence was applied to explore quantum phase transitions in a spin-1 system \cite{Bahmani_2020,ghannadan2021magnetic}. These investigations revealed the presence of two quantum critical points in an Heisenberg antiparallel spin-1 dimer system, accompanied by abrupt changes in thermodynamic quantities at these QCPs.

In this regard, this work analyzes the performance of an anisotropic spin-1 Heisenberg dimer as the working substance of a quantum Stirling machine. By leveraging the interplay between anisotropy, exchange interactions, and external magnetic fields, this study identifies critical conditions where the  engine can achieve optimal thermodynamic performance. We examine the energy spectrum’s dependence on the magnetic field and anisotropy, identifying single, double, and triple degenerate quantum critical points associated with enhanced thermal efficiency and maximum gain. Additionally, we investigate how thermodynamic measures can differentiate between distinct types of QCPs within the system. Therefore, this work not only highlights the feasibility of employing anisotropic Heisenberg dimers as platforms for quantum thermal machines but also lays the groundwork for future experimental validations, particularly in the context of dinuclear metal complex systems. 

\section{Spin model}

Molecular-based spin systems have been increasingly utilized in quantum thermodynamic approaches due to their versatility and well-characterized properties \cite{cruz2022,depaula2024,PhysRevE.110.044135,e22070755,Rojas2024}. Among these materials, the anisotropic spin-1 Heisenberg dimer has emerged as a paradigmatic model for exploring  the impact of its quantum features on their thermodynamic properties \cite{ghannadan2021magnetic,strevcka2005magnetic,STRECKA20083146,makarova2017acs,QE1}.
 
 Thus, by tuning the anisotropy parameters and external magnetic field, the change in the energy spectrum can be explored in a quantum thermodynamic scenario, exploring the impact of anisotropy-induced quantum level-crossings on the operational modes and efficiency of quantum heat engines.
 
\subsection{Hamiltonian Model and Quantum Level Crossing}

In order to analyze the effect of anisotropy-induced quantum level-crossings in a quantum thermodynamic framework, the working medium is modeled as an anisotropic spin-1 Heisenberg dimer given by the following Hamiltonian:
\begin{align}
 \vu{\mathcal{H}} &= 
     {J}\hat{\boldsymbol{S}}_1\cdot \hat{\boldsymbol{S}}_2 + {D}\left[\left( { \hat{S}}_1^z\right)^2+\left( \hat{{S}}_2^z\right)^2\right] +\nonumber \\
     &- {g}_z \mu_z {B}\left( \hat{ {S}}_1^z+ \hat{ {S}}_2^z\right)~. \label{Hamiltonian}
\end{align}
where $\hat{S}_i^j$ denotes the $j$-th Cartesian component $(j \in {x, y, z})$ of the spin operator corresponding to the $i$-th magnetic ion site, $\mu_B$ is the Bohr magneton, and $g_z$ represents the Landé $g$-factor. The exchange coupling constant $J$ sets the characteristic energy scale of the system.
{The term \(D\) represents the single-ion uniaxial anisotropy. For \(D < 0\), the system exhibits easy-axis anisotropy, favoring magnetization along the \(z\)-axis; conversely, for \(D > 0\), the anisotropy becomes easy-plane, promoting magnetization within the \(x\)–\(y\) plane.} Finally, the Zeeman term, $g_z \mu_B B \hat{S}^z_i$, {describes the coupling of each spin to an external magnetic field applied along the \(z\)-axis.} For consistency, we introduce dimensionless variables to express the relevant quantities in {energy units}. The reduced uniaxial anisotropy is defined as $d = D / J$, the reduced magnetic field as $h = g_z \mu_B B / J$, and the reduced temperature as $t = k_B T / J$. The energies $E_n$ for the Hamiltonian {in Eq.~\eqref{Hamiltonian} are listed in} Table~\ref{EnergiesAndKets}, where the detailed definitions of the auxiliary variables are provided  in Eq.~\ref{variables_aux}. It is important to note that the eigenvectors \(\ket{\psi_n}\) are written in terms of the local basis  $\ket{1}= [1,0,0]^T, \ket{0}= [0,1,0]^T,  \ket{-1}= [0,0,1]^T$.

\begin{figure*}  
\centering
\includegraphics[width=1.\textwidth]{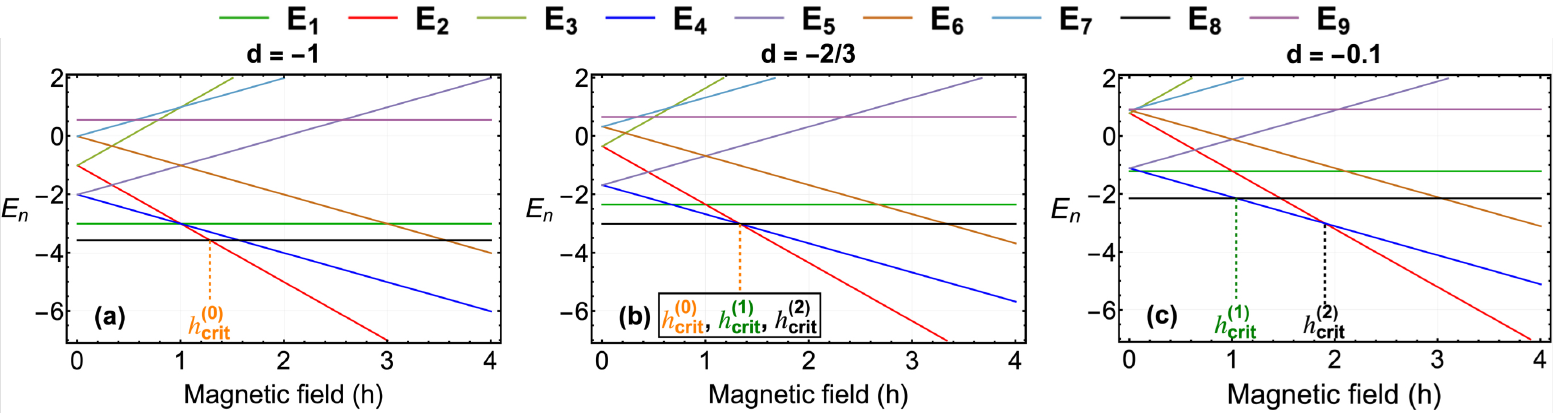}
\caption{Energy spectrum of the antiparallel system as a function of the magnetic field \(h\) for various {negative} anisotropy values \(d\). In panel (\textbf{a}), a QCP, denoted \(h_{\text{crit}}^{(0)}\), emerges where two energy levels cross for \(d < -2/3\). Panel (\textbf{b}) features a triple-degenerate point (TDP) at \(h_{\text{crit}}^{(0)} = h_{\text{crit}}^{(1)} = h_{\text{crit}}^{(2)}\), where three energy levels intersect precisely at \(d = -2/3\). Finally, in panel (\textbf{c}), for \(d > -2/3\), the system exhibits two distinct DDPs, labeled \(h_{\text{crit}}^{(1)}\) and \(h_{\text{crit}}^{(2)}\), respectively.}
\label{PlotEnergies}
\end{figure*}
\renewcommand{\arraystretch}{2}
\begin{table}[h!]
\centering
\caption{Energy \(E_n\) and eigenvectors \(\ket{\psi_n}\)}
\resizebox{.49\textwidth}{!}{%
\begin{tabular}{|c|c|c|}
\hline
\textbf{\(N^o\)} & {Energy} $E_n$ & {Eigenvector} \textbf{\(\ket{\psi_n}\)} \\
\hline
1 & \(J( 2d-1) \) & \(  \qty( \ket{-1, 1} -  \ket{1, -1}) /\sqrt{2} \) \\ \hline
2 & \(J(2d - 2h+ 1  ) \) & \(\ket{1, 1}\) \\ \hline
3 & \(J(2d  + 2h + 1 ) \) & \(\ket{-1, -1}\) \\ \hline
4 & \(J(d - h - 1)\) & \(   \qty(\ket{0, 1} - \ket{1, 0} )/\sqrt{2}  \) \\ \hline
5 & \(J(d + h - 1)\) & \(   \qty(\ket{-1, 0} -\ket{0, -1} ) /\sqrt{2}\) \\ \hline
6 & \(J(d - h + 1)\) & \(   \qty(\ket{1, 0} +   \ket{0, 1}) /\sqrt{2}\) \\ \hline
7 & \(J(d + h + 1)\) & \( \qty( \ket{0, -1} +   \ket{-1, 0})/\sqrt{2} \) \\ \hline
8 & \(-J  (A_1  - 2d+1 )/2\) & 
\(\displaystyle A_3  \qty(\ket{1, -1} +\ket{-1, 1}  )- \frac{A_2 A_3 }{2} \ket{0, 0}  \)  \vspace{.5 mm}\\ \hline
9 & \( J  (A_2)/2\) & 
\(\displaystyle \frac{\sqrt{A_2}}{2 A_1^{1/4}}  \qty(\ket{1, -1}+\ket{-1, 1}) + \frac{2}{A_4} \ket{0, 0}   \) \vspace{.5 mm}\\ 
\hline
\end{tabular}%
}
\label{EnergiesAndKets}
\end{table}
\eq{
\begin{aligned}
    A_1 &= \sqrt{(1 - 2d)^2 + 8},   \\
A_2 &= -1 + 2d + A_1 , \\ 
A_3 &=  \qty({\sqrt{2+ \frac{A_2^2 }{4}}})^{-1}, \\
A_4 &= \sqrt{1 + 8 - A_1 + 2d(-2 + 2d + A_1)}.
\end{aligned} \label{variables_aux}
}

It is important to highlight that the states $\ket{\psi_i}$ $(i = 1 \text{ to } 7)$ have fixed coefficients, making them robust against variations in the exchange interaction, anisotropy, and magnetic field. In contrast, the states $\ket{\psi_8}$ and $\ket{\psi_9}$ are sensitive to changes in the magnetic field and anisotropy, but remain unaffected by the exchange interaction. The dependence of energy levels on the magnetic field leads to crossing points in the energy spectrum for different values of the anisotropy parameter $d$.
For the antiparallel configuration \((J > 0)\), {the system’s behavior depends strongly on the anisotropy parameter \(d\). In particular, one can derive analytical expressions for the critical field values \(h_{\text{crit}}^{(n)}\), which mark the crossings between different quantum energy levels.} These expressions for different critical {magnetic} fields are shown in Eq.~\ref{QCPs}, and can be easily obtained by {explicitly matching} the expressions for the energy levels given {in} Table~\ref{EnergiesAndKets}.

\eq{
\resizebox{.43\textwidth}{!}{$
\begin{aligned}
h_{\text{crit}}^{(0)}: \frac{1}{4}\left(3 + 2d + \sqrt{(1 - 2d)^2 + 8}\right), && d \leq -\frac{2}{3}, \\
\begin{rcases}
h_{\text{crit}}^{(1)}: \dfrac{1}{2}\left(-1 + \sqrt{(1 - 2d)^2 + 8}\right), \\
h_{\text{crit}}^{(2)}: 2 + d  
\end{rcases}, && d \geq -\frac{2}{3}.
\end{aligned}
$}
\label{QCPs}}

For $d < -\frac{2}{3}$, {only a single ground-state level crossing occurs, where the energy \(E_{8}\) transitions to \(E_{2}\)}, governed by the magnetic field at \( h_{\text{crit}}^{(0)} \). Below this field, the entangled state $\ket{\psi_8}$ dominates; once the field exceeds \( h_{\text{crit}}^{(0)} \), the system cross from to the non-entangled state $\ket{\psi_2}$. At the critical anisotropy value $d = -\frac{2}{3}$, the system exhibits a triple-degenerate point (TDP), where the energy levels $E_2 = E_4 = E_8$ coincide. {Despite this degeneracy, on either side of the TDP the system behaves as it does for \(d < -\tfrac{2}{3}\): a single, unique ground-state crossing occurs from \(\ket{\psi_{8}}\) to \(\ket{\psi_{2}}\).} For $d > -\frac{2}{3}$, the system experiences two distinct double degenerate points (DDP). The first {one occurs} at \( h_{\text{crit}}^{(1)} \), where the entangled state $\ket{\psi_8}$ crosses to the entangled state, $\ket{\psi_4}$. The second one arises at \( h_{\text{crit}}^{(2)}\), marking a change from the entangled state $\ket{\psi_4}$ to the pure state $\ket{\psi_2}$. {The primary distinction between these two points lies in which pair of energy levels coincide at each crossing.} These quantum energy levels crossings highlight the intricate interplay between anisotropy \(d\) and magnetic field \(h\) in shaping the quantum phase behavior of the system. This interplay is particularly significant in the context of quantum heat engines, where it is essential to maximize thermodynamic efficiency and approach the Carnot limit. To further elucidate the impact of anisotropy and topology on the properties of the dimer system, the energy levels are shown in Fig.~\ref{PlotEnergies} as functions of the reduced magnetic field \(h\), for various values of the reduced anisotropy \(d\). In Fig.~\ref{PlotEnergies}, the ground state at zero magnetic field is \(\ket{\psi_8}\), as the magnetic field {increases above zero}, the energy levels splits due to the Zeeman effect. 

Specifically, Fig.~\ref{PlotEnergies}(\textbf{a}) illustrates the crossing \(\ket{\psi_8} \rightarrow \ket{\psi_2}\) at the critical magnetic field \(h_{\text{crit}}^{(0)}\), occurring when the anisotropy satisfies the condition \(d < -2/3\). For the critical anisotropy value \(d = -2/3\), depicted in Fig.~\ref{PlotEnergies}(\textbf{b}), three energy levels intersect in the ground state, \(E_2 = E_4 = E_8\), at the critical magnetic field \(h = 4/3\), forming a triple-degenerate point (TDP). 
In {Fig.~\ref{PlotEnergies}(\textbf{c})}, corresponding to higher anisotropy values \((d > -2/3)\), two distinct quantum energy levels crossing emerge at different magnetic field values, \(h_{\text{crit}}^{(1)}\) and \(h_{\text{crit}}^{(2)}\). At the first one, the system {transitions} from \(\ket{\psi_8}\) to \(\ket{\psi_4}\), while at second one, it crosses from \(\ket{\psi_4}\) to \(\ket{\psi_2}\).

The thermodynamic quantities of the system are fully determined by the canonical ensemble. The partition function, constructed from the energies in Table~\ref{EnergiesAndKets}, is given by $
Z = \sum_{n=1}^9 e^{-\frac{\epsilon_{n}}{t}}$,
where {\(t = {k_B T}/{J}\) and $\epsilon_{n}= {E_{n}}/{J}$}. An analytical expression for \(Z\) is obtained as
 \eq{\begin{split}
Z= &e^{-\frac{A_2}{2t}} + e^{\frac{1 - 2d}{t}} +   2 e^{-\frac{1 + 2d}{t}} \cosh\left(\frac{2h}{t}\right) +\\
&+e^{\frac{1 + A_1 - 2d}{2t}}  + 4 \cosh\left(\frac{1}{t}\right) \cosh\left(\frac{h}{t}\right) e^{-\frac{d}{t}} 
.
 \end{split} \label{PartitionFunction}
 }
where $A_1$ and $A_2$ have been previously defined {in Eq.~\eqref{variables_aux}}. 	Using the partition function {defined} in Eq.~\eqref{PartitionFunction}, one can determine, through standard canonical thermodynamic relations, the thermodynamic potentials, such as the Helmholtz free energy $F$ and the internal energy $U$, expressed in terms of the exchange coupling constant $J$. From these, the entropy can also be obtained, naturally expressed in units of the Boltzmann constant $k_B$. Therefore, the relevant thermodynamics quantities for our study is given by the following expressions: 
\eq{
f = F/J&= - t \ln Z, && s = \frac{S}{k_{B}}= -\pdv{f}{t}, && u =\frac{U}{J}= f + t s
}

The thermodynamics of the anisotropic spin-1 dimer could be described using the Gibbs {density} operator for a system in thermal equilibrium, expressed as:
\begin{equation}
    \hat{\rho} = \frac{e^{- \hat{\mathcal{H}}/t}}{\Tr{e^{-  \hat{\mathcal{H}}/t}}} = \sum_{n=1}^9 P_n \ket{\psi_i}\bra{\psi_i},
    \label{rho_total}
\end{equation}
where \(P_n=  {e^{- {\epsilon_n}/t}}/{Z}\) are the Boltzmann occupation probabilities associated with the eigenstates \(\ket{\psi_i}\) in Table~\eqref{EnergiesAndKets} and the canonical partition function {in Eq.}~\eqref{PartitionFunction}.  

This density matrix \(\rho\), as defined in Eq.~\eqref{rho_total}, provides a complete description of the nine-level system comprising two interacting qutrits. The matrix satisfies the fundamental properties of a valid density operator:
\begin{equation}
    \hat{\rho} = \hat{\rho}^\dagger, \quad \Tr{\hat{\rho}} = 1, \quad \text{and} \quad \sum_i \bra{\phi_i}\hat{\rho} \ket{\phi_i} \geq 0,
\end{equation}
ensuring it is Hermitian, normalized, and positive semi-definite.

\subsection{Spin-1 quantum entanglement}
The general definition of the concurrence vector components $C_{\alpha, \beta}$ in terms of the generators of $SO(N_i)$.  \cite{Gconcu8, Bahmani_2020} This serves as the foundation for the framework presented in this section.
\begin{equation}
C_{\alpha,\beta}=\max \left\{0,\lambda_1^{\alpha, \beta}- \sum_{i=2}^{N_1 \times N_2} \lambda_i^{\alpha, \beta} \right\}, \label{concurrence_comps}
\end{equation}
where \(\lambda_i^{\alpha, \beta}\) are the square roots of the eigenvalues of \(\rho \tilde{\rho}\), arranged in descending order. Here, \(\tilde{\rho}\) is defined as \( \left(L_\alpha \otimes L_\beta\right) \rho^\dagger \left(L_\alpha \otimes L_\beta\right)\), where \(\rho^\dagger\) represents the Hermitian conjugate. The generators of \(SO\left(N_1\right)\) and \(SO\left(N_2\right)\) are respectively denoted by \(L_\alpha,\left(\alpha=1, \ldots, \frac{N_1\left(N_1-1\right)}{2}\right)\) and \(L_\beta,\left(\beta=1, \ldots, \frac{N_2\left(N_2-1\right)}{2}\right)\).
In our system, the spin-1 dimer is associated with the generators of \(SO(3)\) \cite{Bahmani_2020}. 
Thus, the generalized concurrence in terms of its components, as given in Eq.~\eqref{concurrence_comps}, becomes:
\begin{equation}
\norm{\mathbf{C}}=\sqrt{\sum_{\alpha, \beta} C_{\alpha, \beta}^2}. \label{general_concurrence}
\end{equation}

The concurrence values described by Eq.~\eqref{general_concurrence} range from zero, representing an unentangled state \((\norm{\mathbf{C}}_{\text{min}}=0)\), to a finite maximum value that depends on the spin dimension of the system. For a system composed of two identical particles with the same $N$-th spin dimension, the maximum concurrence value for a maximally entangled state is given by \( \norm{\mathbf{C}}_{\text{max}}^{SO(N)}=\sqrt{2(N-1) / N}\). For a spin-1 dimer, the maximally entangled state is expressed as: $
\ket{\psi} = \frac{1}{\sqrt{3}} \qty(\ket{1,{-}1} + \ket{0,0} + \ket{{-}1,1}),
$
with a corresponding maximum concurrence of \(\norm{\mathbf{C}}_{\text{max}}^{SO(3)} = 2/\sqrt{3}\) \cite{Bahmani_2020, GConcu9}.
\begin{figure*}
    \centering
\includegraphics[width=.8\linewidth]{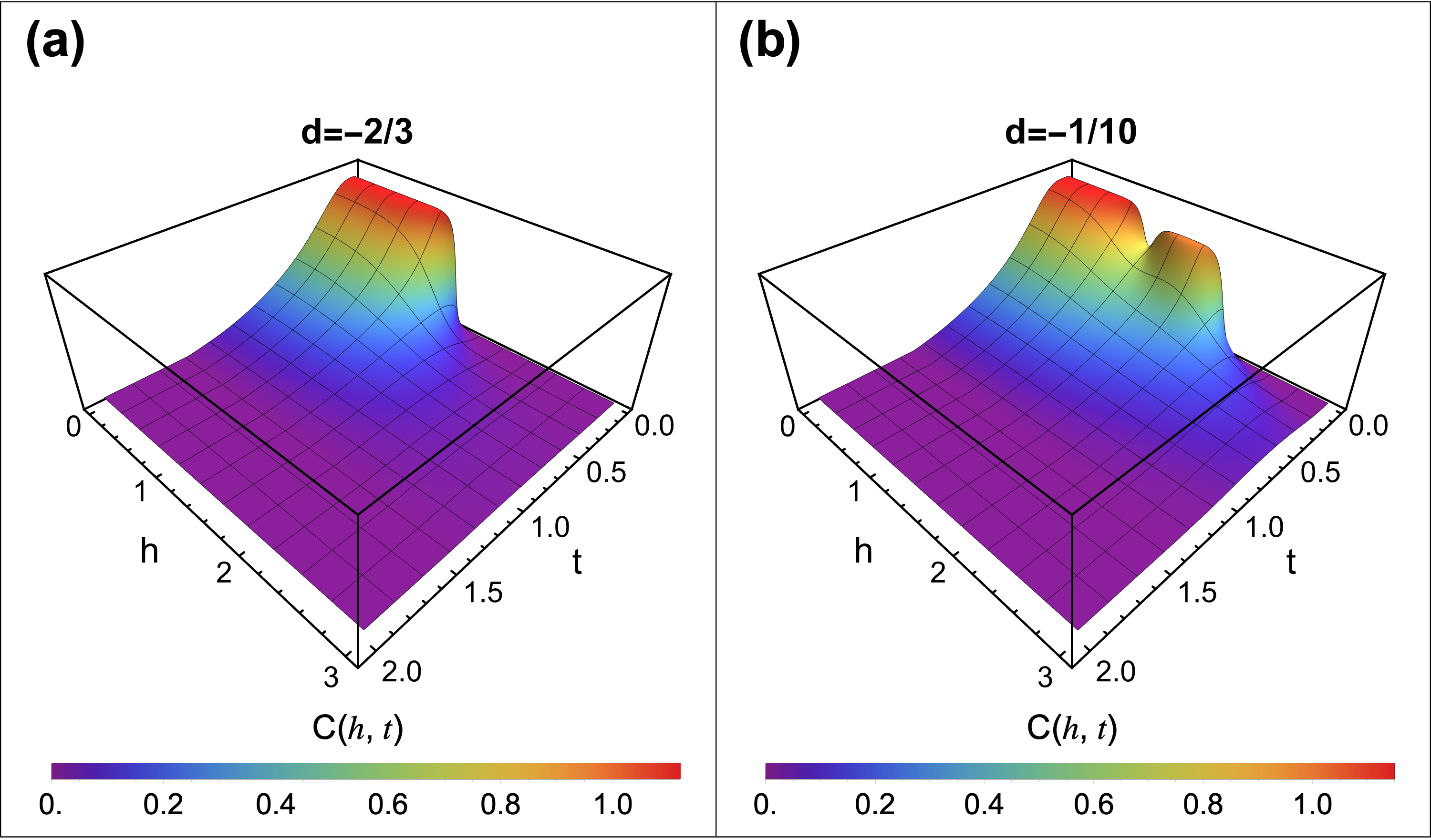}
    \caption{The general spin-1 concurrence $\norm{\mathbf{C}}$ is presented as a function of temperature and magnetic field. Panel $(\mathbf{a})$ corresponds to an anisotropy value of d = -2/3, while panel $( \mathbf{b})$ represents d = -1/10. The maximal concurrence value for both panels is $2/\sqrt{3}$}
    \label{ConcurrenciasPlots}
\end{figure*}
In Fig.~\ref{ConcurrenciasPlots}~$(\mathbf{a})$ and $(\mathbf{b})$, the generalized concurrence is shown as a function of \(T\) and \(h\) for the antiparallel spin-1 dimer. In panel (\textbf{a}), the concurrence for an anisotropy value of \(d = -2/3\) displays an abrupt change at low temperatures, decreasing from the maximum value \(\norm{\mathbf{C}} = 2/\sqrt{3}\) to zero. This change occurs as the ground state evolves from \(\ket{\psi_8}\), which has maximal entanglement under these system parameters, to a state with no entanglement. Exactly at the critical magnetic field \(h_{\text{crit}}^{(0)} = h_{\text{crit}}^{(1)} = h_{\text{crit}}^{(2)} = 4/3\), the system exhibits a triple degeneracy point (TDP), where the states \(\ket{\psi_2}\), \(\ket{\psi_4}\), and \(\ket{\psi_8}\) coexist, and the system's entanglement abruptly drops to zero. For \(h > 4/3\), the ground state changes to the single state \(\ket{\psi_2}\),  characterized by maximal spin alignment {and} fully disentangled.

In Fig.~\ref{ConcurrenciasPlots}~$(\mathbf{b})$, the system is studied with a fixed anisotropy value of \(d = -1/10\). As the temperature approaches absolute zero, the system undergoes two abrupt changes as the magnetic field increases. At zero field, the ground state \(\ket{\psi_8}\) exhibits maximal entanglement. At the first quantum double degenerate point (DDP), \(h_{\text{crit}}^{(1)} \simeq 1.03\), two states, \(\ket{\psi_8}\) and \(\ket{\psi_4}\), coexist, and the concurrence drops to \(\norm{\mathbf{C}} \simeq 0.95\). For magnetic fields between the two DDPs, \(h_{\text{crit}}^{(1)} < h < h_{\text{crit}}^{(2)}\), the concurrence remains constant at \(\norm{\mathbf{C}} = 1\). At the second DDP, \(h = h_{\text{crit}}^{(2)} \simeq 1.9\), the concurrence abruptly drops to zero as the system changes to the maximally aligned spin state.

It is noteworthy that{,} as the anisotropy increases, the entanglement persists over a broader range of magnetic fields. However, in both scenarios, the system quickly loses entanglement due to its pronounced sensitivity to thermal fluctuations. These results are consistent with those reflected by negativity \cite{Bahmani_2020}. {Nevertheless}, neither concurrence nor negativity can distinguish the TDP from a single DDP. In the following sections, we will discuss how the maximum gain (useful work) of the Stirling cycle can distinguish between the different types of QDPs in the system and how these quantum properties can be experimentally measured through thermodynamic quantities at finite temperatures.

\section{Quantum Stirling Cycle}

The modifications in the ground state and the resulting crossings among different quantum levels, which depend on the anisotropy value, significantly affect the population distribution during the magnetic cycles. Consequently, the operational regimes and efficiency of the various heat machines that the system can implement are altered. This aspect will be examined in more detail in this section.

In order to build up any quantum thermodynamic cycles, we start from the First Law of thermodynamics written in the following form \cite{alicki1979quantum}.

\begin{equation} \label{diff_first_Law}
  \dd u = \dbar q + \dbar w = \sum_{n} \epsilon_n \dd P_n + \sum_{n} P_n \dd \epsilon_n,
\end{equation}
where we can see that heat is associated with changes in the occupation probabilities $P_n$ values while work depends on changes in the energy levels $\epsilon_n = E_n/J$.

\begin{figure}[h!]
    \centering
\includegraphics[width=1\linewidth]{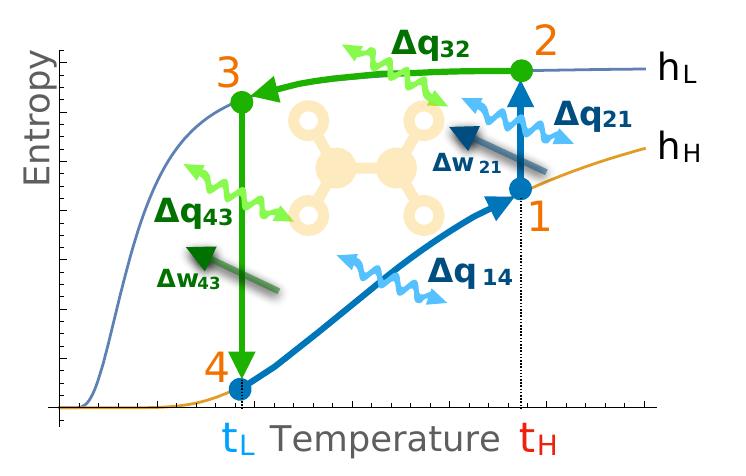}
    \caption{Schematic representation of a magnetic quantum Stirling cycle. Note that work is done on the system ($\Delta w < 0$) during the isothermal ``expansion" ($1 \rightarrow 2$) process, which occurs at thermal equilibrium with the hot thermal bath $t_H$. Conversely, work is performed by the system ($\Delta w > 0$) during the isothermal ''compression" ($3 \rightarrow 4$) process at thermal equilibrium with the cold bath $t_L$. Heat exchange between the system and the reservoirs occurs during all four processes. Note that the sign convention adopted for the work ($\Delta w$) considers the point of view of the magnetic field $h$.}
    \label{Stirling_cycle}
\end{figure}

Quantum Stirling thermal machines have been extensively studied employing numerical modeling using Heisenberg-like spin systems as working substances \cite{huang2014quantum, ghannadan2021magnetic, vargova2021unconventional, cruz2022, araya2023magnetic, PhysRevE.110.044135, Rojas2024, pili2023quantum}. This quantum cycle is a four-stroke closed-loop regenerative heat engine consisting of two (quantum) isothermal processes and two (quantum) isochoric processes, as illustrated in Fig.~\ref{Stirling_cycle}.

During the isothermal steps, the working substance maintains a constant temperature $t$, while both the energy levels $E_n$ and occupation probabilities $P_n$ evolve simultaneously, driven by changes in the applied magnetic field $h$ \cite{Quan:07}. In contrast, during the isochoric strokes, the magnetic field $h$ remains constant, ensuring that no work is performed on or by the system (since $h$ serves as the work parameter). Consequently, only the occupation probabilities $P_n$ change during these steps \cite{kumar2023introduction}.
{Using the First Law as given in Eq.~\eqref{diff_first_Law}, together with standard canonical thermodynamic relations,} one can easily obtain analytical expressions for the heat changes during each one of the processes:

\begin{equation} \label{Q_12}
  \Delta q_{12} =u_2 - u_1 + t_H \ln Z_2 - t_H \ln Z_1,
\end{equation}

\begin{equation} \label{Q_23}
  \Delta q_{23} = u_2 - u_1,
\end{equation}

\begin{equation} \label{Q_34}
  \Delta q_{34} = u_4 - u_3 + t_L \ln Z_4 - t_L \ln Z_3,
\end{equation}

\begin{equation} \label{Q_41}
  \Delta q_{41} = u_4 - u_1,
\end{equation}
The other relevant quantities required to characterize the cycle are the network per cycle $\Delta w$, the heat fluxes $\Delta q_{\text{in}}$ and $\Delta q_{\text{out}}$ and the efficiency $\eta$, given respectively by:
\begin{equation} \label{W_net}
  \Delta w_{23} = w_{12} + w_{34},
\end{equation}
\begin{equation} \label{Q_in}
  \Delta q_{\text{in}} =\Delta q_{12} + \Delta q_{41},
\end{equation}
\begin{equation} \label{Q_out}
  \Delta q_{\text{out}} =\Delta q_{23} + \Delta q_{34},
\end{equation}
\begin{equation} \label{eff}
  \eta = \frac{\Delta w}{\Delta q_{\text{in}}},
\end{equation}
The Clausius formulation of the Second Law of thermodynamics allows for four distinct modes of operation: engine, refrigerator, accelerator, and heater \cite{solfanelli2020nonadiabatic, myers2021quantum}. These modes are classified based on the directions of the heat fluxes, $\Delta q_{\text{in}}$ and $\Delta q_{\text{out}}$, as well as the net work $\Delta w$ per cycle. In this work, we adopt a sign convention from the perspective of the external agent (magnetic field $h$). Accordingly, any energy flux entering the working substance (either as heat or work) is assigned a negative sign (``-"), reflecting the transfer of energy from the field $h$ to the system. Conversely, any energy flux leaving the working substance is assigned a positive sign (``+"). Based on this sign convention, the conditions defining each operation mode are summarized in Table \ref{tab:modes}:\begin{table}[ht]
\centering
\caption{Relationship between the operation modes allowed by the second law of thermodynamics and the signs of the relevant quantities \(\Delta w\), \(\Delta q_{\text{in}}\), and \(\Delta q_{\text{out}}\), following the sign convention from the perspective of the magnetic field \(h\).}
\label{tab:modes}
\begin{tabular}{|l|c|c|c|}
\toprule
\textbf{Operational regime} & \(\Delta w\) & \(\Delta q_{\text{in}}\) & \(\Delta q_{\text{out}}\) \\
\hline
Engine       & \(+\) & \(+\) & \(-\) \\
\hline Refrigerator & \(-\) & \(-\) & \(+\) \\
\hline Accelerator  & \(-\) & \(+\) & \(-\) \\
\hline Heater       & \(-\) & \(-\) & \(-\) \\
\hline 
\end{tabular}
\end{table}

\section{Thermodynamic Implications on the Quantum Level Crossings: Efficiency and Work Output Optimization}

\begin{figure}[h!]
    \centering
    \includegraphics[width=1\linewidth]{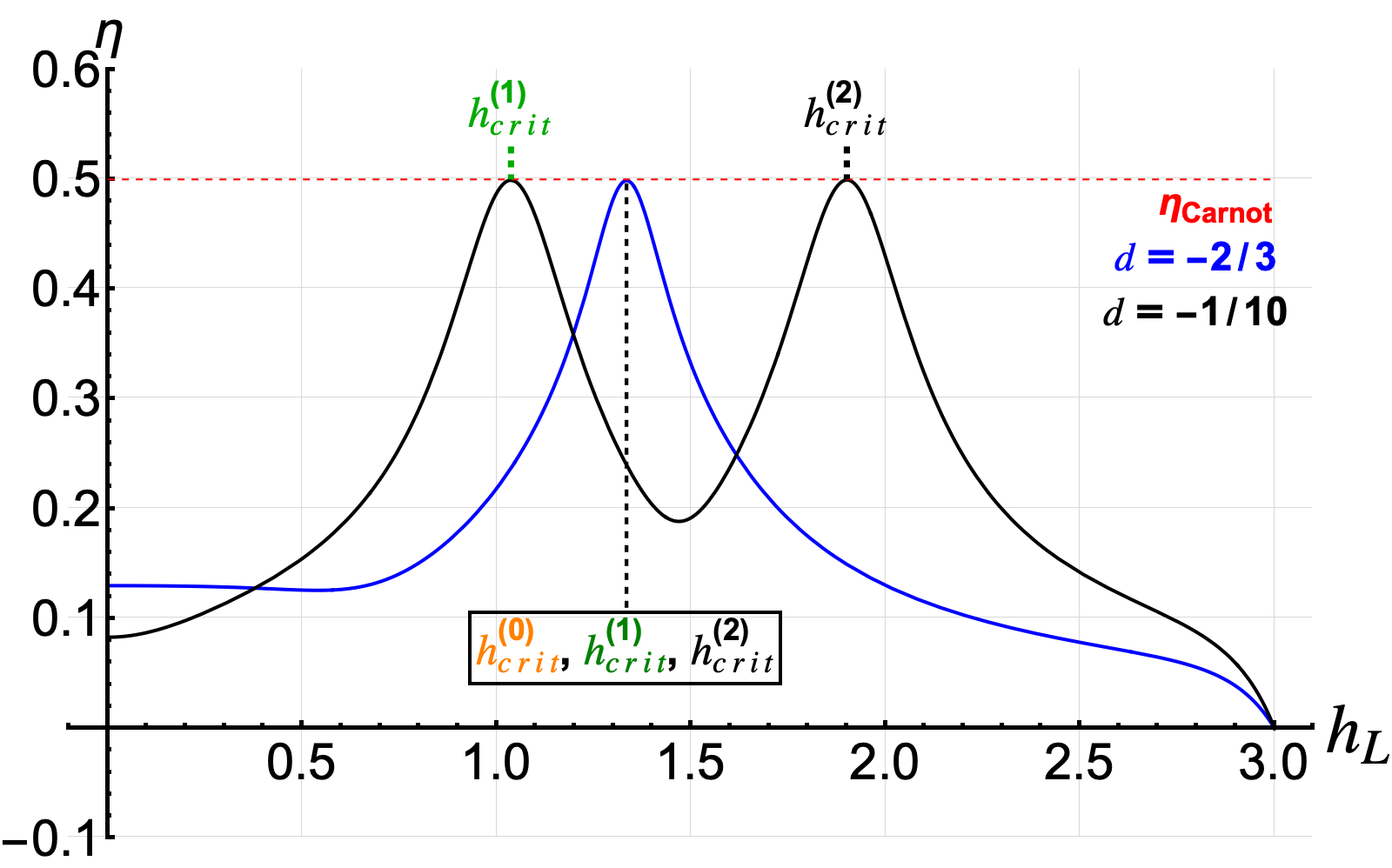}
    \caption{Efficiency \(\eta\) as a function of the low magnetic field \(h_L\) for the antiparallel system. The high magnetic field is fixed at \(h_{\text{H}} = 3.0\), with reservoir temperatures \(t_L = 0.05\) and \(t_H = 0.1\). The dashed red line corresponds to the Carnot efficiency \(\eta_{\text{Carnot}} = 0.5\). {The cycle is examined for two anisotropy values: \(d = -2/3\) (solid blue), which yields a triple-degenerate point (TDP), and \(d = -1/10\) (solid black), associated with two double-degenerate points (DDPs).}}
    \label{Efficiencies}
\end{figure}

In this section, we present results for the antiparallel system ($J>0$). Fig.~\ref{Efficiencies} shows the efficiency $\eta$ as a function of the magnetic field $h_L$, {for two selected values of the uniaxial $z$-axis anisotropy parameter:} $d=-2/3$ (blue solid line) {and} $d=-1/10$ (black solid line). The reservoir temperatures are fixed at $t_L=0.05$ and $t_H=0.1$. We observe that the maximum efficiency occurs precisely at the quantum level crossing points defined in Eq.~\eqref{QCPs}. For $d=-2/3$ (blue curve), the system exhibits a TDP and reaches the Carnot limit at a single magnetic field value, $h_L=4/3$. In contrast, for $d=-1/10$ (black curve) the system presents two quantum critical points (DDPs), {achieving Carnot efficiency at both $h_{\text{crit}}^{(1)}\simeq 1.04$ and $h_{\text{crit}}^{(2)}\simeq 1.90$. 

Although for both values of the anisotropy parameter discussed the system operates entirely as a heat engine under the current conditions, this behavior results solely from the specific choice of temperature and magnetic field parameters. By considering a larger temperature gradient or a narrower spread in the magnetic fields, the system can transition into any of the known operational regimes of thermal machines, as will be explored in greater detail in the following section.}

\begin{figure}
    \centering
    \includegraphics[width=1\linewidth]{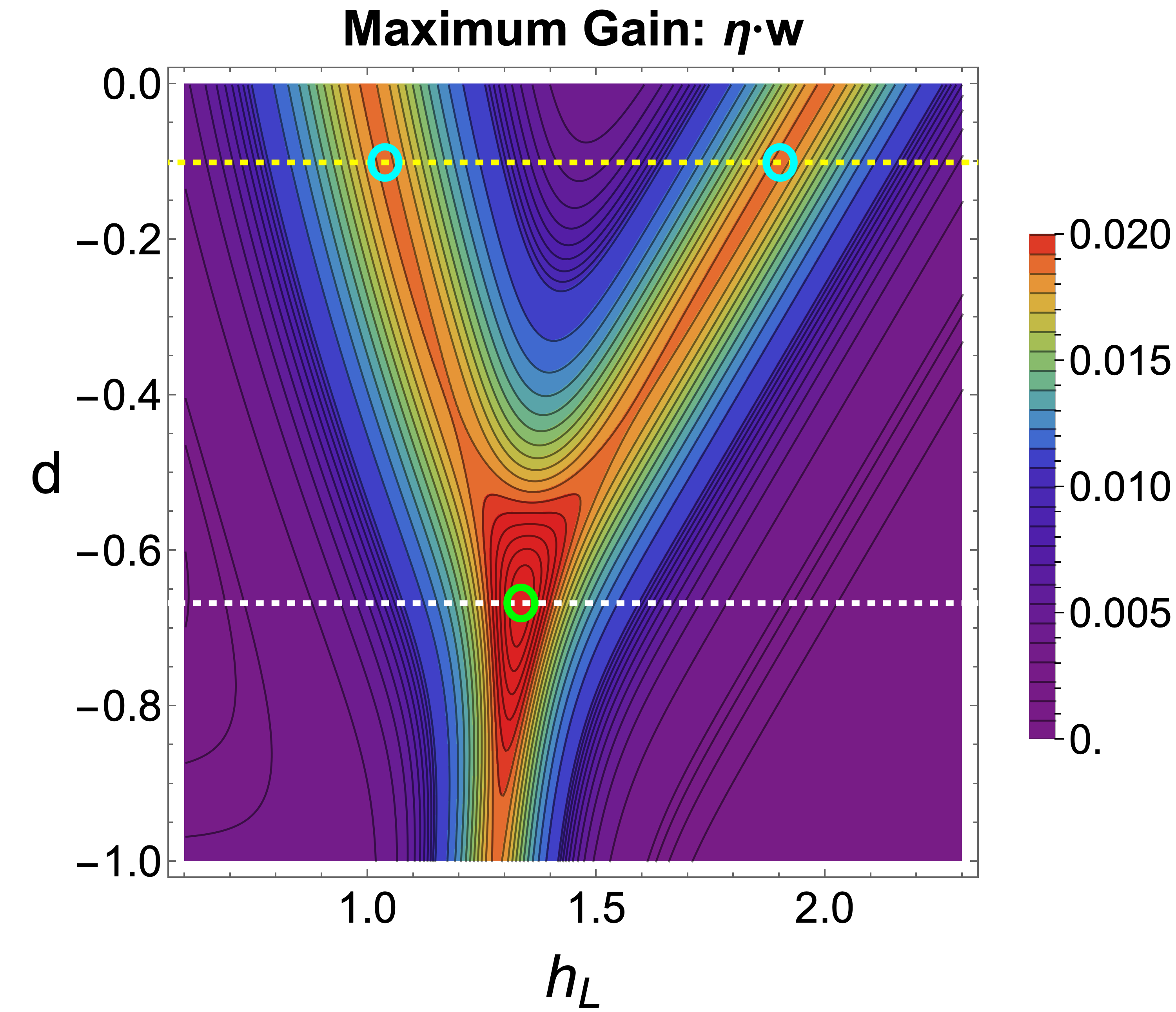}
    \caption{Color map showing the variation of the maximum gain, defined as the product of work and efficiency \((\eta \cdot w)\), for the antiparallel system , as a function of the low magnetic field \(h_L\) ad the uniaxial anisotropy \(d\). The temperatures and high magnetic field are fixed at \(t_L = 0.05\), \(t_H = 0.1\) and $h_H=3.0$. The dashed yellow line {\((d = -1/10)\)} corresponds to the cycle represented by the black solid line in Fig.~\ref{Efficiencies}, with the two cyan circles indicating the Carnot efficiency \(\eta_{\text{Carnot}}\) of the same cycle. The {white} dashed line \((d = -2/3)\) represents the blue solid line in Fig.~\ref{Efficiencies}, and the green circle in it highlights the \(\eta_{\text{Carnot}}\) at the triple-degenerate point given by $ h_{\text{crit}}^{(0)} = h_{\text{crit}}^{(1)} = h_{\text{crit}}^{(2)} = 4/3$.}
    \label{MaximumGain1}
\end{figure}
To evaluate the thermodynamic conditions that maximize the potential for producing useful work, we plot the maximum gain in Fig.~\ref{MaximumGain1}, defined as the product of work and efficiency \((\eta \cdot w)\). {All fixed parameters, such as the temperature reservoirs and the high magnetic field, are set according to Fig.~\ref{Efficiencies}, and the results are analyzed as functions of \(h_{L}\) and \(d\).} The region with the highest useful gain corresponds {precisely to the conditions} at the triple-degenerate point (TDP) \((d = -2/3)\) and $h_L=4/3$. For \(d = -{1}/{10}\), the two double-degenerate points (DDPs) produce identical maximum gains for the cycle{—this statement refers exclusively to the scenario in which DDPs are present}. The plot {indicates} that the maximum-gain regions align exactly with the horizontal positions of the DDPs. Deviations from these critical points lead to a gradual decrease in useful work, underscoring the pivotal role of DDPs in optimizing thermodynamic efficiency {of the quantum heat machine and showing that identical useful work can be obtained at two distinct magnetic field magnitudes: one substantially below the high-field value and the other much closer to it.}
A thermodynamic distinction emerges between the TDP and the DDPs: while both reach Carnot efficiency, the TDP yields a higher useful work output and exhibits superior stability in its surrounding region at finite temperature, maintaining elevated values of useful work despite little variations in \(d\) and \(h_L\).

\begin{figure}[h!]
    \centering
    \includegraphics[width=.82\linewidth]{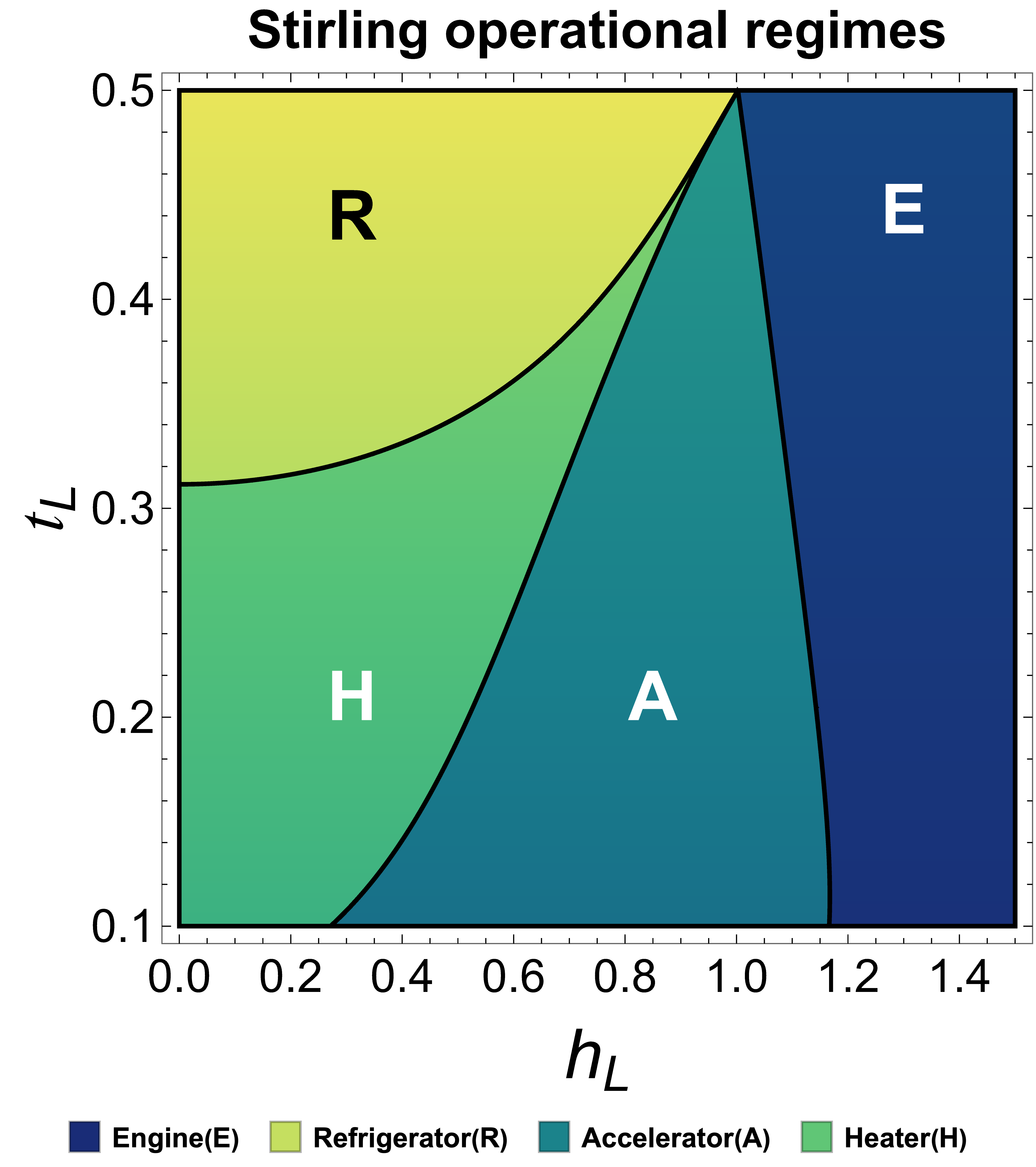}
    \caption{Stirling cycle operational regimes of the isotropic antiparallel system for different values of the lower magnetic field \(h_L\) and low temperature \(t_L\), with fixed parameters: anisotropy {\(d = -0.4\)}, high temperature {\(t_H = 0.5\)}, and high magnetic field {\(h_{\text{H}} = 1.5\)}. The system exhibits all four operational regimes.}
    \label{stirling_operation_regimes1}
\end{figure}
{To examine how these distinct behaviors manifest under different thermal conditions, we now turn to the temperature constraints of the Stirling cycle. All operating temperatures are kept below} the critical temperature \(t_{\text{crit}} = 0.1\) (in natural units of the model), as the cycle does not function entirely as a thermal engine for \(t > t_{\text{crit}}\) while maintaining the same magnetic field {magnitudes}. To characterize the various Stirling machine behaviors exhibited for different \(t_L\) and \(h_L\), Fig.~\ref{stirling_operation_regimes1} shows the system's operational regimes under fixed parameters: {\(t_H = 0.5\), \(h_{\text{H}} = 2.0\) and $d=-0.4$}. The system demonstrates all four operational regimes.

{When the cycle is operated at a fixed low temperature, $ t_{L}\approx 0.32$, while varying the magnetic field $h_{L}$, the system transitions across all four operational regimes. Conversely, if the magnetic field is held constant at $h_{L}\approx 0.4$ and $t_{L}$ is varied, the system goes through three distinct regimes but never functions as a heat engine. However, when \(h_{L}\) is close to the high fixed-field value (\(h_{L}\approx 1.3\)), the cycle functions exclusively as a heat engine as \(t_{L}\) varies.}

\section{Conclusions and Perspectives}

In summary, this work presented an anisotropic spin-1 Heisenberg dimer model as a robust theoretical framework for exploring the interplay between quantum thermodynamic effects and induced quantum level crossings. A systematic investigation of energy levels and the concurrence has revealed two double-degenerate points (DDPs) a triple-degenerate point (TCP) for the system considering antiparallel alignment. Multiple energy levels converge at the ground state at specific values of reduced anisotropy and reduced magnetic field, resulting in a discontinuity in the Concurrence. In a quantum thermodynamic approach, under these conditions, the system is predicted to achieve Carnot efficiency and maximum power output. In addition, the versatility of this quantum thermal machine is highlighted by its ability to operate in all thermodynamically allowed modes given by the Clausius formulation of the second law of thermodynamics. Thus, the spin-1 Heisenberg dimer model presented in this work not only advances our understanding of induced quantum level crossings on quantum thermodynamics but also provides practical insights for designing and controlling quantum thermal machines.

\begin{acknowledgments}
{B.C, F.J.P., and P.V. acknowledge financial support from ANID
Fondecyt grant no. 1240582 and 1250173}. B.C acknowledge PUCV and UTFSM. {B.C. acknowledges the support of ANID Becas/Doctorado Nacional 21250015}. V.G.P thanks Rio de Janeiro State Research Support Foundation (FAPERJ) for the financial support. C. Cruz thanks the Fundação de Amparo à Pesquisa do Estado da Bahia - FAPESB for its financial support (grant numbers APP0041/2023 and PPP0006/2024). This study was financed in part by the Coordenacao de Aperfeicoamento de Pessoal de Nivel Superior – Brasil (CAPES) – Finance Code 001. M.S.R. thanks FAPERJ and belongs to the INCT of Refrigeração e Termofísica, funding by CNPq by grant number 404023/2019-3. CICECO-Aveiro Institute of Materials, is supported by UIDB/50011/2020, UIDP/50011/2020 \& LA/P/0006/2020, financed by national funds through the FCT/MCTES (PIDDAC).  
\end{acknowledgments}

{\section*{Conﬂict of Interest}
The authors declare no conﬂict of interest.
\section*{Data Availability Statement}
Data sharing is not applicable to this article as no new data were created or analyzed in this study.
\section*{Keywords}
{Quantum Thermodynamics, Spin-1 Heisenberg dimer, Quantum level crossing, Magnetic anisotropy,  Quantum Stirling engine }
}
\appendix



%

\end{document}